\documentstyle[12pt,ssi,amssymb]{article}
\input epsf
\epsfverbosetrue

\begin{document}

\title{Recent Results from KamLAND}

\author{Jason Detwiler\\
Department of Physics \\
Stanford University, Stanford, CA 94305 \\[0.4cm]
Representing the KamLAND Collaboration }

\maketitle
\begin{abstract}
The Kamioka Liquid-scintillator Anti-Neutrino Detector (KamLAND)
has detected for the first time the disappearance of electron
antineutrinos from a terrestrial source at the 99.95\%
C.L.\cite{KLPRL} Interpreted in terms of neutrino oscillations
$\bar{\nu}_e \leftrightarrow \bar{\nu}_x$, the best fit to the
KamLAND data gives a mixing angle $\sin^2 2 \theta = 1.0$ and a
mass-squared difference $\Delta m^2 = 6.9 \times 10^{-5}$ eV$^2$,
in excellent agreement with the Large Mixing Angle (LMA) solution
to the "solar neutrino problem".\cite{solarprob} Assuming CPT
invariance, this result excludes other solutions to the solar
neutrino problem at $>$ 99.95\% C.L.
\end{abstract}

\section{Reactor Antineutrino Experiments and Neutrino Oscillation}

Nuclear reactors emit a calculable flux of electron antineutrinos
($\bar{\nu}_e$s) in all directions. For standard particle
propagation, one expects a detector located a distance $L$ from
the reactor to measure a flux that decreases as $1/L^2$. But if
$\bar{\nu}_e$s are massive, they may "oscillate" into undetectable
flavors on the way to the detector, leading to an apparent
"disappearance" of the $\bar{\nu}_e$s.

Neutrinos are produced and detected in weak interactions, which
couple to the weak eigenstates $\nu_l$, where $l = e, \mu, \tau$.
For massive neutrinos, the weak eigenstates may be expressed as a
linear combination of three mass eigenstates $\nu_i$, $i = 1, 2,
3$, with mass $m_i$:
\begin{equation}
\nu_l = \sum_i U_{li} \nu_i. \label{eSuperpos}
\end{equation}
$U_{li}$ is a $3 \times 3$ unitary "mixing" matrix and is
analogous to the CKM matrix in the quark sector. As a neutrino
propagates through vacuum with momentum $p_\nu$, the phase of each
mass eigenstate will change at different rates according to
\begin{equation}
\nu_l(t) = \sum_{i} U_{li} e^{-i E_i t} \nu_i,
\label{ePropogation}
\end{equation}
where $E_i \approx p_\nu + \frac{m_i^2}{2 p_\nu}$. At times $t
> 0$ the neutrino will be in a superposition of the weak
eigenstates. The probability of detecting flavor $\nu_{l'}$ at a
distance $L \approx t$ from the source is then
\begin{equation}
P(\nu_l \rightarrow \nu_{l'}) = Re \sum_{i,j} U_{li} U_{il'}^\dag
U_{lj} U_{jl'}^\dag e^{i \frac{m_i^2 - m_j^2}{2 p_\nu}L}.
\label{e3FlavProb}
\end{equation}

In neutrino oscillation experiments, it is common to simplify to
two neutrino flavors, say $\nu_e$ and $\nu_\mu$, in which case
$U_{li}$ is parameterized by a single mixing angle $\theta$. In
this picture, the probability of a neutrino emitted as $\nu_e$ to
be detected as $\nu_e$ is then
\begin{equation}
P \left( \nu_e \rightarrow \nu_e \right) = 1 - \sin^2 2 \theta
\sin^2 \left( \frac{\pi L}{L_{osc}} \right) \label{e2FlavProb}
\end{equation}
where $L_{osc} = \frac{4 \pi p_\nu}{\Delta m^2}$. Here, $\Delta
m^2 = \left| m_1^2 - m_2^2 \right|$ is taken to be the
mass-squared difference between two "effective" mass eigenstates
relevant to $\nu_e \leftrightarrow \nu_\mu$ mixing. Note that
$P\left( \nu_e \rightarrow \nu_e \right)$ is a sinusoidal function
of L, hence the name "neutrino oscillation".

The situation becomes slightly more complicated for neutrinos
propagating through matter. As first recognized by
Wolfenstein\cite{Wolf}, Mikheyev, and Smirnov\cite{MS}, while all
three weak states participate in neutral-current interactions with
normal matter, only electron neutrinos have additional
charge-current interactions with electrons in the material being
traversed. This results in an effective index-of-refraction for
$\nu_e$ different from that for $\nu_{\mu, \tau}$ by
\begin{equation}
\Delta n = \frac{2 \pi N_e}{p_\nu^2} f(0), \label{eRefInd}
\end{equation}
where $N_e$ is the number-density of electrons in the material,
and $f(0)$ is the forward scattering amplitude. $\Delta n$ changes
the phase of the $\nu_e$-component relative to the other
components by $2 \pi$ over a distance $L_0$ given by
\begin{equation}
L_0 = \frac{2 \pi}{\sqrt{2} G_F N_e}, \label{eRefInd}
\end{equation}
where $G_F$ is the Fermi constant. This phenomenon, called the MSW
effect, alters the two flavor oscillation probability as follows:
\begin{equation}
P \left( \nu_e \rightarrow \nu_e \right) = 1 - \sin^2 \left( 2
\theta_m \right) \sin^2 \left( \frac{\pi L}{L_{m}} \right)
\label{eMSW1}
\end{equation}

\begin{equation}
\tan 2 \theta_m \equiv \tan 2 \theta \left( 1 +
\frac{L_{osc}}{L_0} \sec 2 \theta \right) \label{eMSW2}
\end{equation}

\begin{equation}
L_m \equiv L_{osc} \left[ 1 + \left( \frac{L_{osc}}{L_0} \right)^2
+ \frac{2 L_{osc}}{L_0} \cos 2 \theta \right]^{- \frac{1}{2}}
\label{eMSW3}
\end{equation}

For neutrinos created in and propagating out of the sun, $L_0
\approx 200$ km $\ll R_{sun}$, and matter effects are significant.
The result is that $\nu_e$ disappearance experiments that use the
sun as a source, such as Homestake\cite{Homestake},
GALLEX\cite{GALLEX}, SAGE\cite{SAGE}, Kamiokande\cite{Kamiokande},
Super-Kamiokande\cite{SuperKSolar}, and SNO\cite{SNO},
individually allow values of $\tan^2 \theta$ and $\Delta m^2$ over
many orders of magnitude. The overlap of allowed regions from
multiple experiments with different thresholds leaves a few small
patches, with values around $\tan^2 \theta \approx 1$ and $\Delta
m^2 \approx 10^{-4}$ eV$^2$ highly favored. This region is called
the "Large Mixing Angle" (LMA) MSW solution. Prior to KamLAND,
another patch at lower $\Delta m^2 \approx 10^{-7}$ eV$^2$, known
as the "LOW" MSW solution, survived at the 99.73\% C.L.\cite{SNO}

For neutrino experiments with artificial sources on the earth,
matter effects are typically much less significant. This is
because for matter with density similar to rock $L_0 \approx 10^4$
km, larger than the radius of the earth. Thus for these sources
the modifications to the oscillation parameters can be viewed as
perturbations on the vacuum parameters, and for short enough
oscillation lengths can be neglected.

Artificial $\nu$ sources for oscillation experiments typically
take one of two forms: neutrino beams or nuclear reactors. These
two sources comprise a highly complementary experimental program
in neutrino oscillation. Neutrino beams in general require very
long baselines for good $\Delta m^2$ sensitivity, but the
collimation of these sources lessens the severity of this obstacle
considerably. Moreover, beams produce higher energy neutrinos and
multiple flavors, allowing for not only disappearance but also
appearance measurements. Reactors, on the other hand, are a
non-collimated source, and produce exclusively electron
(anti)neutrinos, restricting the experimenter to
disappearance-only measurements and thus giving limited
sensitivity to $\sin^2 2 \theta$. However, due to the lower
energies of the neutrinos (1-10 MeV), reactor experiments are
sensitive to very small $\Delta m^2$.

Note that both neutrino beams and reactors are "laboratory-style"
experiment, in which both the source and the detector are
controlled. The physics of neutrino propagation can then be
cleanly tested separately from the mechanisms involved in the
production of the neutrinos.

\section{Calculating Reactor Antineutrino Fluxes}

More than 99.9\% of $\bar{\nu}_e$s emitted by nuclear reactors are
produced in the decay chains of the fission products of only four
isotopes: $^{235}$U, $^{238}$U, $^{239}$Pu, and $^{241}$Pu. The
number of fissions of each of these isotopes over the data-taking
period is estimated from a Monte Carlo simulation of the reactor
core that tracks fuel burn-up and U/Pu production over reactor
fuel cycles. The performance of such simulations has been verified
by comparing the simulated fuel composition at the end of a fuel
cycle with measurements of isotopic abundances in the actual spent
fuel rods.

The inputs to the simulation include the initial fuel composition
and periodic measurements of the secondary calorimetric power, the
pressure and flow rate in the primary cooling system, and various
other operational parameters. However, the results of the
simulations depend very weakly on most of the inputs. Accuracy is
required only for the fuel composition and the power, on which the
fission rates depend linearly. The power is measured rather
precisely: reactors are regulated to operate at about a standard
deviation or so below their rated power outputs. The economic
incentive to produce as much power as possible pushes this error
to be small; uncertainties $< 1\%$ are common.

The emitted $\bar{\nu}_e$ spectrum is then calculated by summing
the spectrum emitted by each isotope weighted by the number of
fissions of that isotope during the data taking period. For
$^{235}$U, $^{239}$Pu, and $^{241}$Pu, $\bar{\nu}_e$ spectra are
derived from $\beta$-spectra
measurements.\cite{Hahn}$^,$\cite{Schreckenbach} Extracting the
$\bar{\nu}_e$ spectra from the $\beta$-spectra is non-trivial; it
involves complex bookkeeping of the decay chains for each daughter
nucleus and summing the contributions. The spectra are normalized
to the number of $\bar{\nu}_e$s emitted per fission, and have
uncertainties of a few percent. For $^{238}$U, no measurements are
available, so we must rely on a calculation\cite{238U} of the
$\bar{\nu}_e$ spectrum. The uncertainty in the calculation is
$\sim$10\%, but since fissions from $^{238}$U make up only
$\sim$10\% of the signal, this results in a 1\% uncertainty in the
full $\bar{\nu}_e$ flux.

Reactor experiments typically detect $\bar{\nu}_e$s via
inverse-$\beta$-decay, $\bar{\nu}_e + p \rightarrow n + e^+$,
because the 2-particle final state distinguishes the $\bar{\nu}_e$
from most other particles and dramatically reduces backgrounds.
The positron carries away almost all of the energy of the incident
neutrino, so that the detected positron spectrum is to a good
approximation the incident neutrino spectrum folded with the
detection cross section and shifted by a constant energy. The
cross section has a threshold of 1.8 MeV, and has been
calculated\cite{xsection} with uncertainties on the level of a
percent.

The most recent generation of reactor experiments, Palo
Verde\cite{PaloVerde} and Chooz\cite{Chooz}, showed that
uncertainties on the level of a few percent can be achieved for
these calculations. The fact that $\bar{\nu}_e$ disappearance has
not been detected in reactor experiments with baselines up to
$\sim$1 km\cite{PaloVerde}$^-$\cite{Bugey} allows one to view
these experiments as a verification the reactor flux calculations
to within a few percent over most of reactor neutrino spectrum.

\section{The KamLAND Experiment}

KamLAND uses the entire Japanese nuclear power industry as a 180
GW$_{th}$ long-baseline source. The experiment is located
underneath Mt. Ikenoyama in Gifu prefecture in central Japan. 80\%
of the $\bar{\nu}_e$ flux comes from reactors at baselines of
140-210 km. This range of baselines makes KamLAND particularly
sensitive to values of $\Delta m^2$ corresponding to the LMA MSW
solution to the solar neutrino problem. Matter effects in rock can
be neglected at this distance scale.

The results presented here are for 145.1 live-days of data taking
between March and September 2002. Calculating the incident flux at
KamLAND during this period requires summing the flux from every
reactor in Japan. Reactor data, including instantaneous power and
fuel burn-up, is provided by the Japanese power companies. The
flux from reactors in Korea and the rest of the world is estimated
from reported power generation, and makes up only a few percent of
the overall flux.

As stated earlier, KamLAND detects $\bar{\nu}_e$s via
inverse-$\beta$-decay, $\bar{\nu}_e + p \rightarrow n + e^+$, with
the energy of the positron $E_{e^+}$ related to the energy of the
incident $\bar{\nu}_e$ by essentially a constant offset of 1.8
MeV. The positron quickly deposits this energy in KamLAND's liquid
scintillator before annihilating with an electron, giving a prompt
energy deposit of $E_\nu - 0.8$ MeV. Meanwhile, the neutron
thermalizes and eventually captures on a proton, releasing a 2.2
MeV gamma in the process. The characteristic capture time of
$\sim$ 210 $\mu$s between the prompt and delayed events provides a
powerful background reduction.

\begin{figure}
\centerline{\epsfxsize=5.0in \epsfbox{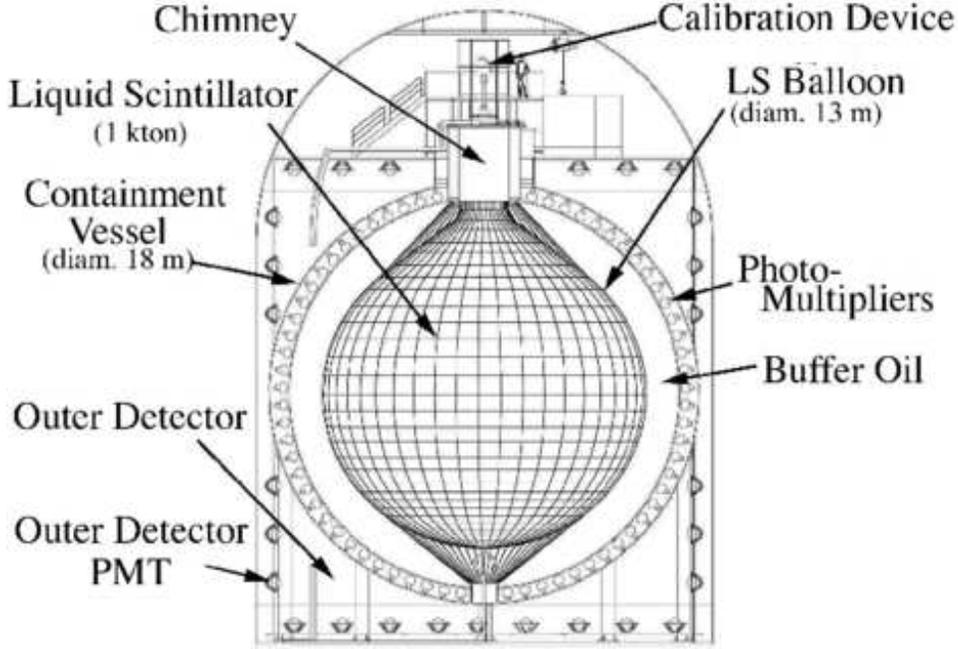}}
\caption[]{Schematic diagram of the KamLAND detector.}
\label{fig:detector}
\end{figure}

KamLAND's liquid scintillator is composed of 80\% dodecane, 20\%
pseudocumene, and 1.52 g/liter of PPO. One kton of scintillator is
held inside a 6.5 m radius transparent balloon made of a 135
$\mu$m thick sandwich of nylon and ethylene vinyl alcohol
copolymer films. 1879 17- and 20-inch photomultiplier tubes (PMTs)
mounted on the inner surface of an 18 m stainless steel sphere
view events inside the scintillator. For the results reported
here, only the 1325 17-inch tubes are used, corresponding to 22\%
coverage. A buffer of mineral oil between the PMTs and the balloon
provide buoyancy for the balloon and shielding from radioactivity
in the PMT glass. Outside the steel sphere is a water cherenkov
outer detector (OD) serving as a muon veto. A chimney and deck at
the top of the detector provide access for calibration devices to
be deployed along the z-axis. A schematic of the detector is given
in Figure~\ref{fig:detector}.

An event is recorded whenever 200 or more PMTs report a signal
within 125 ns. A delayed-event window is then opened for 1 ms with
a lower threshold of 120 tubes. The pulses are digitized by Analog
Transient Waveform Digitizers (ATWDs). There are two ATWDs per PMT
to reduce dead time during digitization.

\begin{figure}
\centerline{\epsfxsize=3.0in \epsfbox{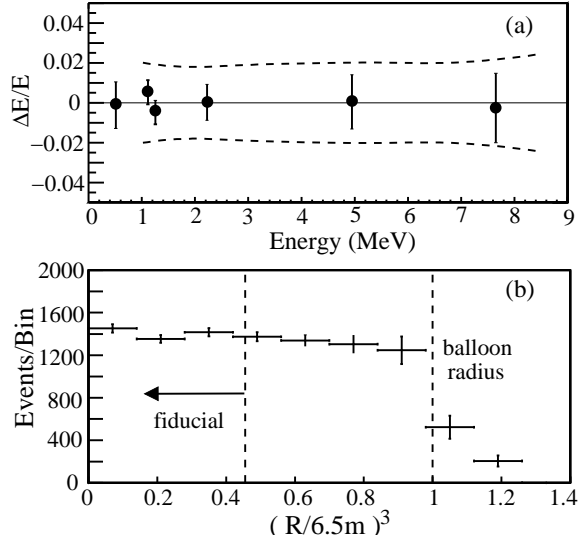}}
\caption[]{(a) Fractional error in energy reconstruction relative
to the known source energies. The dashed line shows the systematic
uncertainty in the energy scale. (b) Distribution in R$^3$ of 2.2
MeV capture gammas from neutrons following muons. The ratio of
events occurring inside the fiducial volume cut gives the
fractional size of the fiducial volume.} \label{fig:eAndR}
\end{figure}

The visible energy and position of each event is reconstructed
from the collected charge and the hit pattern. The reconstruction
algorithms are developed and tested using uniformly distributed
spallation products following muons and using calibration data
along the z-axis with $^{68}$Ge, $^{60}$Co, and $^{65}$Zn gamma
sources, and an AmBe neutron/gamma source. The calibration data
are also used to estimate the conversion from visible energy to
kinematic energy, taking into account particle-dependent
non-linear effects due to quenching and cherenkov light
production. The systematics on the energy scale is shown in the
upper panel of Figure~\ref{fig:eAndR}.

Antineutrino events are selected according to the following
criteria: time correlation 0.5 $\mu$s $< \Delta t <$ 660 $\mu$s;
vertex correlation $\Delta R <$ 1.6 m; prompt event energy $E_p
>$ 2.6 MeV; delayed event energy 1.8 MeV $< E_d <$ 2.6 MeV;
spherical fiducial volume $R <$ 5 m; cylindrical radius $\rho >$
1.2 m. The cut on $E_p$ eliminates $\bar{\nu}_e$ backgrounds from
geological U/Th sources. The cylindrical radius cut removes
backgrounds from thermometers deployed along the z-axis to monitor
the scintillator. The total efficiency of these cuts was
determined to be 78.3 $\pm$ 1.6\%, and the residual accidental
backgrounds, estimated using an off-coincidence time correlation
window 20 $\mu$s $< \Delta t <$ 20 s, is negligible at $< 10^{-5}$
events per day. The size of the fiducial volume is determined by
counting the ratio of uniformly distributed spallation products
that pass the fiducial volume cut. The distribution for spallation
neutron capture gammas is given in the bottom panel of
Figure~\ref{fig:eAndR}.

The correlated backgrounds passing the $\bar{\nu}_e$ cuts are
produced by cosmogenic spallation. An overburden of 2700 m.w.e.
reduces the muon rate to $\sim$ 0.3 Hz. Muons are identified by
their large energy deposit and outer detector activity, and the
muon track is reconstructed from the PMT hit pattern. A 2 ms veto
following muons eliminates backgrounds due to spallation neutrons
and short lifetime spallation products. Long lifetime spallation
products that mimic the $\bar{\nu}_e$ signal, such as $^8$He and
$^9$Li, are removed by a 2 second veto in a 3 m radius around the
muon track, plus a 2 second veto over the entire volume following
muons with very high energy deposits ($\gtrsim$3 GeV). The
residual correlated backgrounds, estimated from spallation event
studies, time-since-last-muon distributions, and, for fast
neutrons following muons not tagged by the OD, Monte Carlo
simulations, is 1 $\pm$ 1 event in 145.1 days.

\begin{figure}
\centerline{\epsfxsize=4.0in \epsfbox{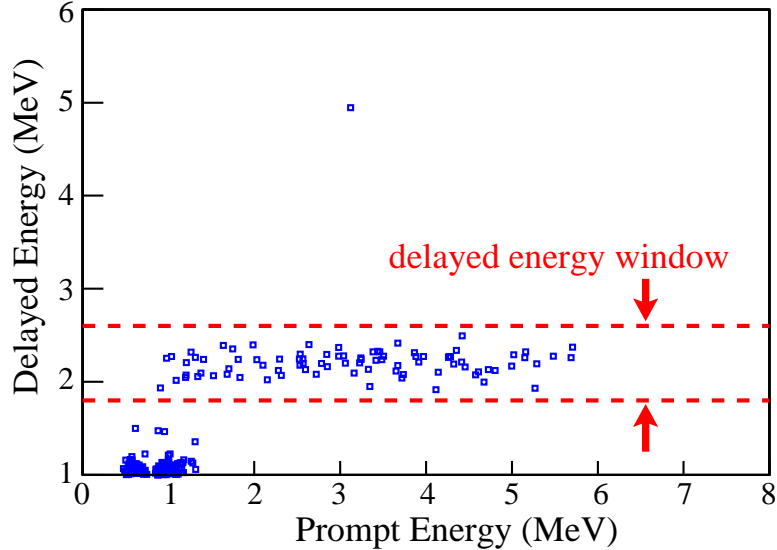}}
\caption[]{Prompt and delayed energies of $\bar{\nu}_e$
candidates.} \label{fig:eped}
\end{figure}

Figure~\ref{fig:eped} shows the distribution of the $\bar{\nu}_e$
candidates' prompt and delayed energies. The delayed 2.2 MeV gamma
cleanly separates $\bar{\nu}_e$ from the accidental coincidences
clustered at small energies. The one event with $E_d \approx$ 5
MeV is consistent with neutron capture on carbon; such events are
not considered in this analysis at a small cost in efficiency with
negligible uncertainty.

The systematic uncertainties are listed in Table~\ref{tab:sys}.
The largest contribution to the systematics comes from the
estimate of the fiducial mass ratio. The total uncertainty is
6.4\%.

\begin{table}
\caption{Estimated systematic uncertainties (\%). }
\label{tab:sys}
\begin{center}
\begin{tabular}{llll}
\hline \hline
Total LS mass       & 2.1  &\ \ \ \ Reactor power                            & 2.0 \\
Fiducial mass ratio & 4.1  &\ \ \ \ Fuel composition                         & 1.0 \\
Energy threshold    & 2.1  &\ \ \ \ Time lag                                 & 0.28   \\
Efficiency of cuts  & 2.1  &\ \ \ \ $\nu$ spectra \cite{Hahn}$^-$\cite{238U} & 2.5 \\
Live time           & 0.07 &\ \ \ \ Cross section \cite{xsection}            & 0.2 \\
\hline
Total systematic error & & & 6.4\% \\
\hline \hline
\end{tabular}
\end{center}
\end{table}

\begin{figure}
\centerline{\epsfxsize=5.0in \epsfbox{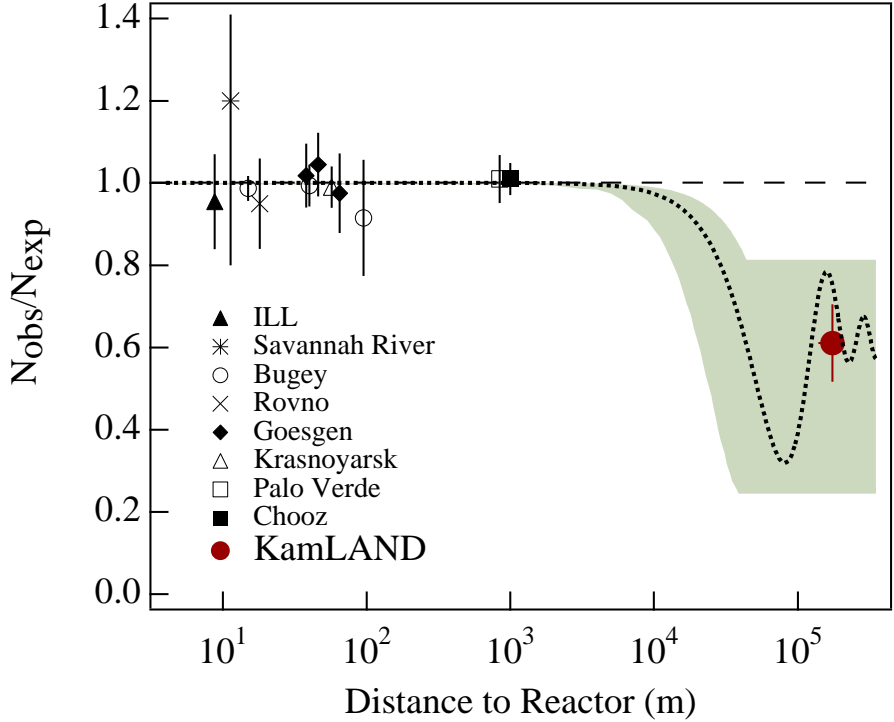}}
\caption[]{$N_{obs}/N_{no~osc}$ for reactor $\bar{\nu}_e$
experiments.} \label{fig:reacratios}
\end{figure}

For no oscillations, in 145.1 live-days we expect 86.8 $\pm$ 5.6
events with 1 $\pm$ 1 of these coming from backgrounds. The
observed number of events is 54, giving a ratio $\frac{N_{obs} -
N_{BG}}{N_{no~osc}} = 0.611 \pm$ 0.085 (stat) $\pm$ 0.041 (syst).
This ratio is plotted vs. baseline for KamLAND and previous
reactor experiments in Figure~\ref{fig:reacratios}. The shaded
region corresponds to the LMA MSW solution to the solar neutrino
problem, with which the reactor experiments are in excellent
agreement. "Standard" $\bar{\nu}_e$ propagation is excluded at $>$
99.95\% C.L.

\begin{figure}
\centerline{\epsfxsize=4.0in \epsfbox{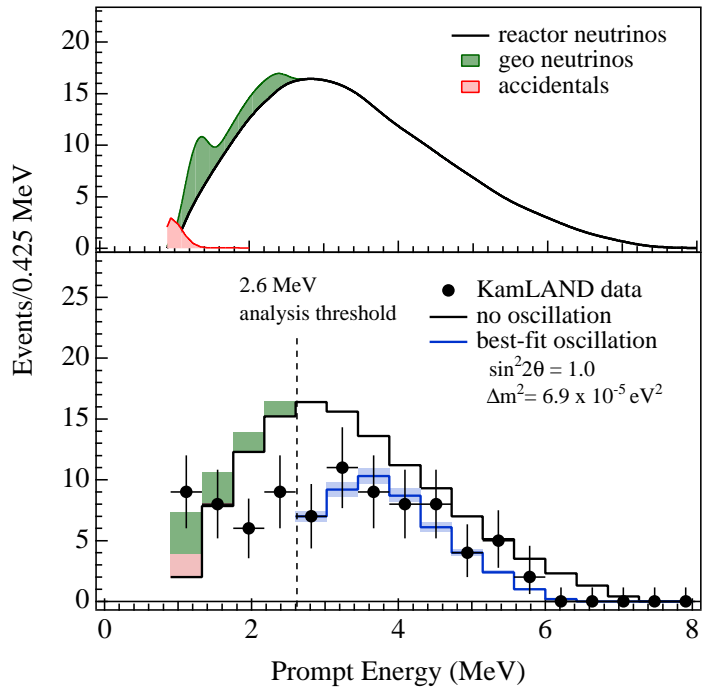}}
\caption[]{Prompt energy spectrum.} \label{fig:spectrum}
\end{figure}

\begin{figure}
\centerline{\epsfxsize=4.0in \epsfbox{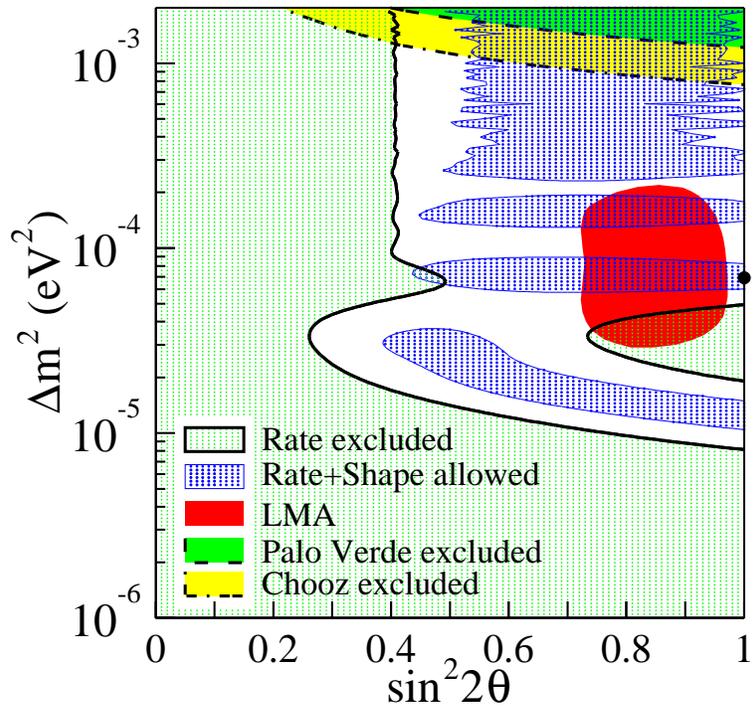}}
\caption[]{95\% inclusion region from KamLAND rate + shape.}
\label{fig:shapeinc}
\end{figure}

The positron energy spectrum is histogrammed in
Figure~\ref{fig:spectrum}. An unbinned maximum likelihood fit was
performed and included terms for changes in the overall rate and
shape of the spectrum at different points in $\Delta m^2 - \sin^2
2 \theta$ parameter space. The best fit solution is plotted as the
shaded histogram, and corresponds to $\Delta m^2 = 6.9 \times
10^{-5}$ eV$^2$ and $\sin^2 2 \theta = 1.0$. The regions in
parameter space allowed at 95\% C.L. are drawn in
Figure~\ref{fig:shapeinc}. Also shown is the exclusion region in
$\Delta m^2 - \sin^2 2 \theta$ parameter space based on the
KamLAND rate alone. Assuming CPT invariance (i.e. that $\nu$ and
$\bar{\nu}$ masses are identical), all neutrino oscillation
solutions to the solar neutrino problem are excluded except LMA.
The KamLAND result divides LMA into two regions of higher and
lower $\Delta m^2$. The sensitivity in $\sin^2 2 \theta$ is rather
poor. Values of $\Delta m^2$ between the inclusion regions in
general predict large distortions in the energy spectrum that were
not observed and are hence disfavored. Constant suppression of the
non-oscillated spectrum is consistent with the data at 53\% C.L.
The inclusion region extends to high $\Delta m^2$ where spectral
distortions can not be distinguished with current statistics and
energy resolution. An upper limit can be placed on $\Delta m^2$
using the non-observation of $\bar{\nu}_e$ disappearance by Palo
Verde and Chooz.

\section{Future of the KamLAND Reactor Measurement}

\begin{figure}
\centerline{\epsfxsize=5.0in \epsfbox{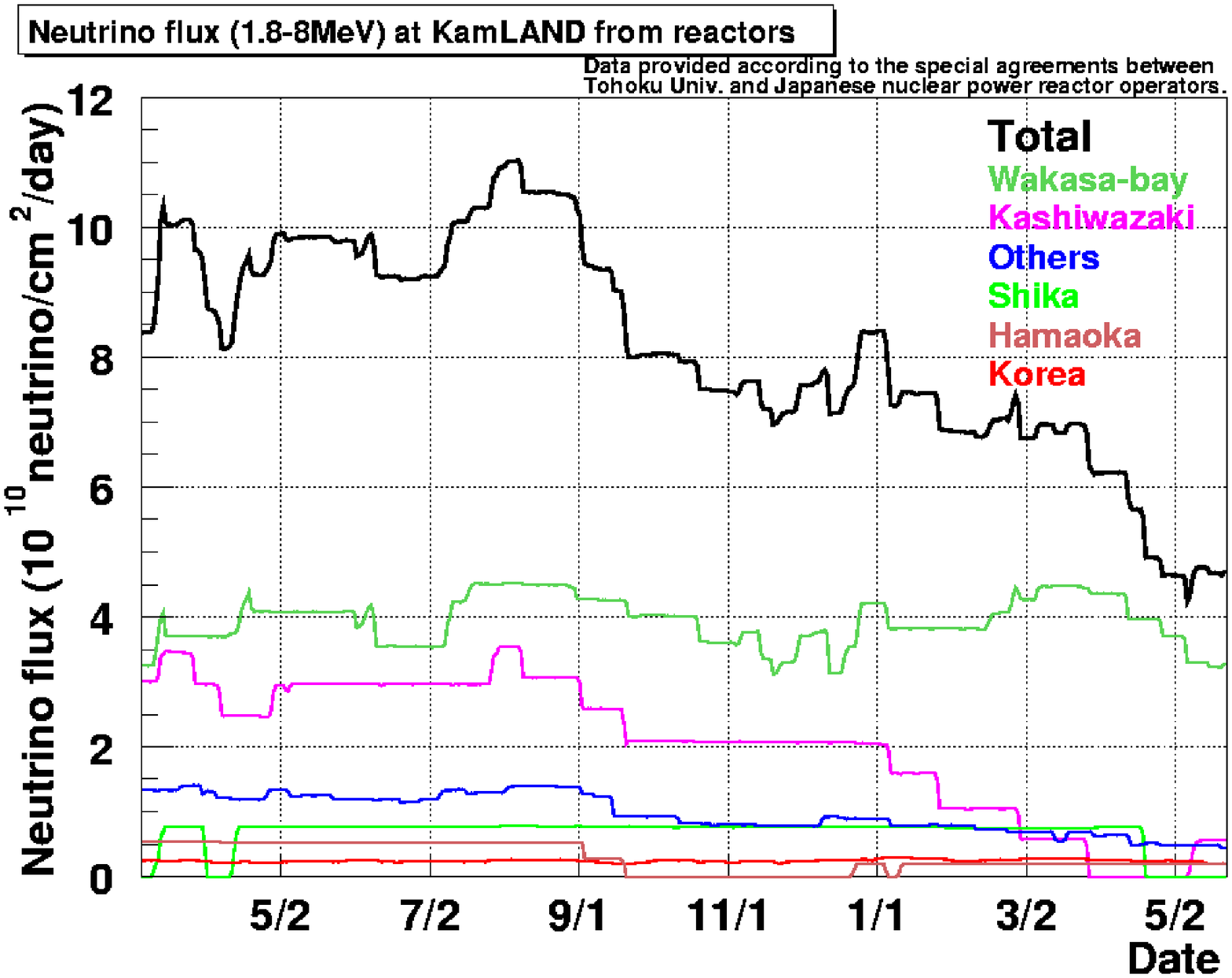}}
\caption[]{Time variation of the $\bar{\nu}_e$ flux at KamLAND.}
\label{fig:futflux}
\end{figure}

Figure~\ref{fig:futflux} shows the $\bar{\nu}_e$ flux at KamLAND
from March 2002 through May 2003. In early 2003, Japanese
utilities powered down their reactors for inspections and
maintenance, resulting in a drop in the non-oscillated flux at
KamLAND by about a factor of two. Such a drastic time variation
will provide a more precise estimate of backgrounds, and depending
on the "true" values of $\Delta m^2$ and $\sin^2 2 \theta$, may
also provide a time-varying spectral shape that will further
restrict the inclusion regions in parameter space.

\begin{figure}
\centerline{\epsfxsize=3.5in \epsfbox{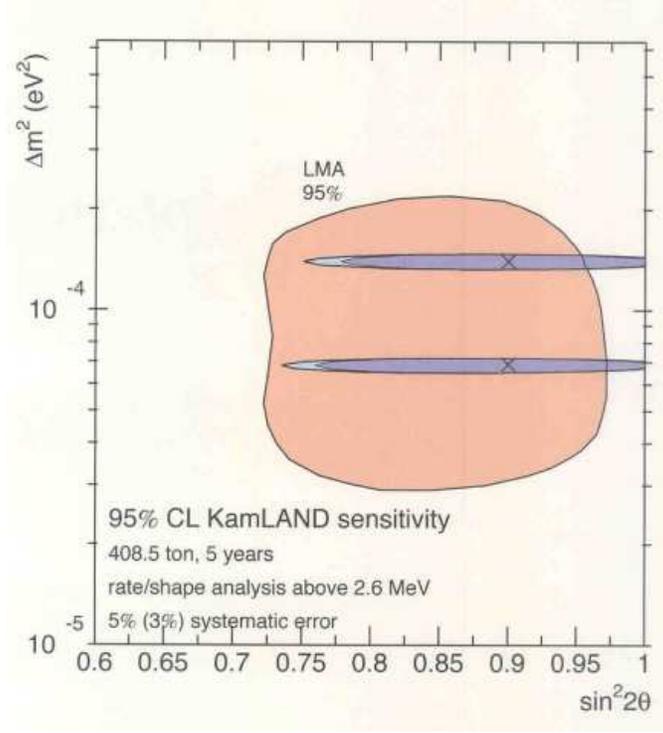}}
\caption[]{Expected KamLAND sensitivity for 5 years of data.}
\label{fig:futsens}
\end{figure}

The expected sensitivity of KamLAND for 5 years of data taking is
shown in Figure~\ref{fig:futsens} for two sets of parameters
expected from KamLAND's first results. KamLAND stands to provide
exquisite sensitivity in $\Delta m^2$. However, further
restrictions on $\sin^2 2 \theta$ will probably have to come from
future solar neutrino experimental results.

\section{Future of Reactor Experiments}

With the latest generation of solar and reactor experiments dawns
a new age in precision neutrino oscillation physics. The next step
in this exciting field will be to improve on current measurements
and to measure some of the remaining unknown parameters in full
three-flavor mixing. In addition to the three masses $m_i$, there
are 6 free parameters in the mixing matrix $U_{li}$. It is
possible to parameterize $U_{li}$ as follows:
\begin{equation}
U = U_{LMA} \times U_{atmospheric} \times U_{13} \times U_{\beta
\beta}. \label{e3FlavMixMat}
\end{equation}
\begin{eqnarray}
U_{LMA} & = & \left( \begin{array}{ccc}
\cos \theta_{12} & \sin \theta_{12} & 0 \\
-\sin \theta_{12} & \cos \theta_{12} & 0 \\
0 & 0 & 1
\end{array} \right) \\
U_{atmospheric} & = & \left( \begin{array}{ccc}
1 & 0 & 0 \\
0 & \cos \theta_{23} & \sin \theta_{23} \\
0 & -\sin \theta_{23} & \cos \theta_{23}
\end{array} \right) \\
U_{13} & = & \left( \begin{array}{ccc}
\cos \theta_{13} & 0 & e^{-i \delta_{CP}} \sin \theta_{13} \\
0 & 1 & 0 \\
-e^{-i \delta_{CP}} \sin \theta_{13} & 0 & \cos \theta_{13}
\end{array} \right) \\
U_{\beta \beta} & = & \left( \begin{array}{ccc}
1 & 0 & 0 \\
0 & e^{-i \frac{\alpha}{2}} & 0 \\
0 & 0 & e^{-i \frac{\alpha}{2} + i \beta}
\end{array} \right)
\end{eqnarray}
$U_{atmospheric}$ and the value of $\Delta m^2_{23}$ has been
measured by the Super-Kamiokande experiment.\cite{SuperKAtm}
$U_{\beta \beta}$ and the overall mass scale will require
kinematical and neutrino-less double-$\beta$ decay measurements.
KamLAND is playing a central role in determining $U_{LMA}$ as well
as the value of $\Delta m^2_{12}$. Reactor experiments stand to
play another role in neutrino oscillation, this time in the
measurement of $U_{13}$.

Because $\Delta m^2_{12} \ll \Delta m^2_{23}$, it must be the case
that $\Delta m^2_{13} \approx \Delta m^2_{23}$. A new generation
of reactor experiments has been proposed to search for
$\bar{\nu}_e$ disappearance at baselines of ~1 km corresponding to
this value of $\Delta m^2$. To improve on the mixing angle
sensitivity achieved by Palo Verde and Chooz, proposals for
reactor $\theta_{13}$ experiments include a large detector to
reduce the statistical error, and also a second detector
positioned very close ($\sim$100 m) to the reactor. The near
detector would precisely measure the incident flux, allowing many
of the flux calculation systematics to drop out. This also
requires that the detectors be made identical and/or moveable.
Sensitivity down to $\sin^2 2 \theta_{13} \approx 10^{-2}$ seems
within grasp. Such experiments were first discussed by Mikaelyan
and Sinev \cite{Mikaelyan}; for a comprehensive list of references
on this topic, please see the web site compiled by
Heeger.\cite{KarstenSite}

\section{Conclusion}

KamLAND has observed, for the first time, disappearance of
electron antineutrinos in a laboratory-style experiment. Assuming
CPT invariance, this result excludes solar neutrino oscillation
solutions except LMA at $>$ 99.95\% C.L. Recall that the LMA
solution to the solar neutrino problem is a region in oscillation
parameter space allowed from the overlap of many experimental
results using different techniques and with different thresholds.
Moreover, LMA is an MSW solution for neutrinos, in which matter
effects inside the sun drive the oscillations. It is significant
that KamLAND, which uses yet another detection technique and is
sensitive to {\it vacuum} oscillations of {\it anti}neutrinos,
gives oscillation parameters in agreement with LMA. These
experiments are different in so many aspects, yet the physics of
neutrino oscillation ties them together in a beautifully
consistent theoretical framework.

\section{Acknowledgements}

The KamLAND experiment is supported by the Center of Excellence
program of the Japanese Ministry of Education, Culture, Sports,
Science and Technology, and the United States Department of
Energy. The reactor data are provided courtesy of the following
electric associations in Japan: Hokkaido, Tohoku, Tokyo, Hokuriku,
Chubu, Kansai, Chugoku, Shikoku, and Kyushu Electric Power
Companies, Japan Atomic Power Co., and Japan Nuclear Cycle
Development Institute. Kamioka Mining and Smelting Company
provided services for activities in the mine.

\end{document}